# Conceptualization and Quantitative study of Aesthetic and Affective Perception of Pictures in Physics Education

Tatjana Zähringer[1*], Raimund Girwidz[1], Andreas Müller[2]

## Abstract

Beyond their well-known instructional functions, pictures in physics education serve many affective functions such as attracting attention, sometimes creating fascination, and fostering engagement with the depicted physics content. Assuming a pivotal role of these functions, it becomes evident that better understanding of the aesthetic perception and a research-based educational use of aesthetic pictures is necessary. Specifically, prior research suggests that perceived aesthetic and affective attractiveness fosters enjoyment in interacting with pictures and increases the readiness to engage with the depicted physics content.

Based on this, the present paper provides three contributions: Firstly, it provides a conceptualization and set of research-based criteria for a selection of pictures that are supposed to be perceived as aesthetic by learners. These criteria are derived from research about the psychology of aesthetic perception, among others as applied to visuals in information and communication technology. Additionally, content-related criteria are derived from research and practice in physics education. Following these criteria, aesthetic pictures related to a given curricular content can be selected.

Secondly, the paper applies the derived criteria to the selection of aesthetic pictures in the field of geometrical optics. It then delves into an evaluation of students' aesthetic and affective perception of the selected pictures. For this, an instrument consisting of the aesthetic and affective perception of science-related pictures was developed and validated (aesthetic perception: $\alpha_C = 0.87$ [0.85, 0.89]; affective perception: $\alpha_C = 0.82$ [0.80, 0.85]).

Thirdly, we present an approach to combine the decorative and instructional functions of pictures in tasks, and a classroom study on the aesthetic and affective perception of the selected pictures in these tasks. Students at junior high school level ($N = 118$) worked on physics tasks with aesthetic pictures (AP, selected according to the derived criteria) or with pictures showing classroom experiments (CEP), covering the same content. In a crossover design, the aesthetic and affective perception of pictures in both categories were then compared. Results revealed a significantly better evaluation of the selected aesthetic pictures in tasks, with large effect sizes compared to pictures showing classroom experiments in tasks covering the same physics content (AP vs. CEP, aesthetic and affective perception: $d = 1.05 - 1.56$ and $0.85 - 1.48$, respectively).

We conclude that the criteria developed and investigated in this study are useful for selecting pictures that evoke a noticeable positive aesthetic and affective perception in learners. This provides a basis for further leveraging their educational potential to create fascination and engagement in science education.

[1]Ludwig-Maximilians-Universität München, [2]University of Geneva, Switzerland
* Zähringer née Lamparter; tatjana.lamparter@physik.uni-muenchen.de

# I. Introduction

Print and electronic media are replete with colorful visualizations. Investigations into their role in physics education identified various functions, including showing phenomena, recalling information, organizing content and motivating readers and learners [1]. A well-known example is the visual appeal of astronomic pictures [2]. Pictures in textbooks in mathematics [3] and physics, chemistry, or biology [4] comprise a wide variety of visualization types, in particular colorful photographs.

Photographs and other pictures that are visually appealing, and not primarily intended to depict concrete physics contexts or be incorporated in exercises or tasks, are referred to as decorative pictures [5]. In work by Lindner and in similar studies [5 - 8], effects of decorative pictures on affective and cognitive outcomes were investigated. Our study intends to implement photographs that are decorative and instructional at the same time.

In order to use the educational potential of visual material in physics exercises and tasks, a key question is which criteria can be used to select pictures that are aesthetically and affectively appealing. A research-based, easy-to-implement list of criteria would be desirable for better understanding and practical use of aesthetic effects in physics education.

Criteria to classify physics related pictures as aesthetically pleasing are provided by psychology of visual perception [9], as well as by physics education research about affective aspects [10]. However, to the best of our knowledge, no studies were carried out to investigate theoretically and experimentally the characterization, selection and application of aesthetic pictures in physics education.

This paper first gives an overview of the literature about the functions of pictures in STEM education, and a conceptualization and set of research-based criteria for the selection of pictures perceived as aesthetic by learners (Sec. II). It then applies these criteria to the selection of aesthetic pictures for physics teaching[1], presents an approach to combine the decorative and instructional functions of pictures in tasks, and the design, instruments, and analysis of a classroom study on the aesthetic and affective perception of the selected pictures in these tasks (Sec. III). The results of this study are presented in Sec. IV, and the findings, limitations and implications for research and teaching practice are discussed in Sec. V.

# II. Research background and objectives of the study

### A. Functions of pictures in STEM education: Instructional and decorative pictures

A current distinction in research about pictures in education in general [7, 9] and science education in particular [5, 11, 12] is between instructional and decorative (decorational) pictures. This distinction is based on their main function: providing information and instructional content on the one hand, and aesthetic experience and emotional valence on the other hand.

A wealth of research on educational effects is available for instructional pictures in general [13 - 15], specifically for science education [12, 16, 17]; and in particular in connection with multiple representations [18 - 20]. In current developments, particularly in physics education, there is an ongoing integration of diverse media forms such as animations, augmented or virtual reality, educational games, and many more. However, these novel integrations still fundamentally rely on (dynamic) pictures as foundational instructional elements [21, 22, and literature therein].

Much less is known about the educational effects of decorative pictures, even though, historically, the use of pictures, emphasizing their aesthetic effects, dates back centuries. A famous example is Comenius's 'Visible World in Pictures' (1658), which aimed to foster children's curiosity, joy of learning, and ultimately, comprehension by an educational compendium of pictures intended to be aesthetic and instructional at the same time [23].

The term 'decorative (decorational) pictures' in its modern sense was coined by Levin, as appealing or interesting pictures 'without a clear instructional purpose' [24, 25]. For a long time, there was a widespread belief that decorative pictures distract from the learning content and are therefore counter-productive, with only instructional pictures being considered relevant for learning and teaching. Already in the work by Levin, based on reviews of prior research, it was concluded that decorative pictures do not facilitate learning from text, as they do not follow the 'principle that pictures in text must, in some sense, be related to that text' [24]. This is consistent with general findings that affective arousal may reduce the working memory capacity available for cognitive processing and thus have a negative impact on learning [26 - 28].

---

[1] Several of the following considerations will also apply to STEM education in general.

A highly influential study strongly supporting this point of view was carried out by Harp & Mayer [25]. They understand decorative pictures as 'seductive details' and conclude from their findings that such details 'hurt student learning of a scientific explanation' (both for seductive pictures and text). Note that in their choice of decorative pictures, these were irrelevant to the specific purpose of the learning text (e.g., the causes of lightning), but not to the topic in general (i.e. lightning and its effects). In fact, from a teacher's point of view, their pictures (e.g., a lightning stroke on an airplane) were indeed interesting and apt to promote a motivating context for the given learning content.

This perspective on decorative pictures as 'seductive details', impairing learning 'by an increase of learning-irrelevant cognitive processes' [7], was later integrated into the cognitive theory of multimedia learning (CTML) developed by Mayer and colleagues [29]. In that theory, drawing from cognitive load theory, a coherence principle for successful instructional design is stated. This principle requires omitting all elements – such as decorative pictures – that are not related to the learning content itself, as they create extraneous cognitive load detrimental to learning. Summarizing the literature (covering effect sizes from 11 comparisons), Mayer concludes that 'there is strong and consistent evidence for the coherence principle', i.e., learning is considerably enhanced (median effect size $d = 1.32$) when extraneous material is not included (both for classical and multimedia, [29]. These findings about a detrimental 'seductive detail effect' on learning were confirmed in a meta-analysis by Rey, with large effect sizes for *not* using decorative pictures on retention ($d = 0.95$) and transfer ($d = 0.83$) [30].

In the last decade, this thinking tradition 'decorative = seductive = detrimental' has been increasingly challenged both on grounds of empirical evidence and of theoretical progress. Lenzner et al. showed that in a lower secondary level classroom on ray optics, decorative pictures in the learning material can induce better mood, alertness, and calmness, but show little or no distracting effects [6]. Moreover, it was found that decorative pictures can reduce the perceived difficulty of the learning content. Plass et al. discuss how aesthetically pleasing pictures can facilitate comprehension [31]. Plass & Kaplan note that designs with different aesthetics initiate different emotions and these emotions can influence the perception of learners and their cognitive processing [32]. They also note, that other researchers found that multimedia elements like the visual design 'resulted in positive user perceptions about learning' [32]. These findings are consistent with earlier research suggesting that positive effective states in general can have a positive impact on learning [33, 34]. Conversely, in a lower secondary general science classroom, Lindner found that decorative pictures in test items had no substantial detrimental effects on performance [5]. The current view is that empirical evidence in this field is inconclusive [35, 36], with dozens of studies finding inhibitory effects of decorative pictures [29, 37] or, conversely, no inhibitory effects [37].

From a theoretical perspective, a review on the role of emotions in cognitive load theory emphasizes affective processing, particularly students' emotional experiences during learning and other cognitive processes [36]. From their review, the authors conclude 'that a comprehensive understanding of cognitive processing requires the consideration of affective factors', i.e., of feelings or emotions experienced during the learning process [36, p. 340]. In particular, they consider how to incorporate affective factors into processing models of multimedia learning and draw on the Cognitive-Affective Theory of Learning with Media (CATLM), which expanded CTML by considering affective mediation [38 - 40]. According to CATLM, motivation, i.e., the processes that direct and sustain a person's long-term attitude and short-term behavior towards learning, mediates learning by influencing cognitive engagement of the learners. Specifically, CATLM emphasizes the importance of affective factors as critical component for the allocation of working memory resources to learning contents and tasks. Consequently, functions such as eye-catching, holding attention, and establishing contextualization became crucial. In a general perspective, CATLM treats learning 'as a result of the interaction between attitudes, emotions, and knowledge' [36]. This interaction determines the 'regulation of cognitive processes', and thus strongly influence principles of instructional design which is both affectively and cognitively stimulating and effective.

Specifically, a recent strand of research providing new perspectives and interesting results is about 'related', decorative pictures, i.e., visually appealing pictures that do not have an explanatory function but are purposefully related to the text and content being conveyed, serving to create a visual context for it [7, 35, 41]. Indeed, in classical studies on decorative pictures, these were adjunct to the text and learning content, just juxtaposed to it without providing any connection [25].

Studies have been carried out on related decorative pictures with an auxiliary role, where they do not attempt to draw more attention to the picture than to the main content of the text but might improve learning through cognitive activation processes initiated by the created visually appealing context; this has indeed been confirmed by recent findings [41].

In the present work, we will investigate an even stronger form of related decorative pictures, namely a form of pictures integrating decorative and instructional functions, specifically aesthetic pictures in tasks (see Sec. III C. for description and example). As a basis for this, we will next review research that establishes criteria for selecting pictures perceived as aesthetic by learners, complemented by affective and practical criteria for the application in physics tasks.

### B. Aesthetic, affective, and practical criteria for picture selection

While aesthetic preferences may be subjective, several basic principles and criteria for them have been shown to be relevant by theories of the psychology of perception of aesthetics and art. We will follow Leder et al., who, drawing on foundational work by Berlyne and related studies, developed a model of aesthetic experience based on color, visual complexity, contrast and Gestalt laws as key features [42, 43]. In the following sections, we will consider these features as criteria for selecting visually appealing pictures.

Color as primary stimulus plays a pivotal role, as Leder states consistent with many other authors. For instance, Deng emphasizes its significance in enhancing perceived aesthetics and interest for photographs [44]. Moshagen & Thielsch also confirm the prominence of color as an aspect of perceiving the aesthetics of websites [45]. In recent work, Leder identifies a preference for more vivid colors, as long as natural tones are preserved [46]. Joye et al. demonstrate that photographs with the context 'nature' are perceived more aesthetic than pictures with urban context [47]. As one main reason they indicate that nature scenes in contrast to urban scenes are intrinsically colorful.

As for visual complexity, Leder notes a preference for moderate levels, consistent with Berlyne, too low or high complexity levels being perceived as boring or overwhelming [42, 43]. Similarly, Schmidt and Wolff found that websites with medium complexity received the highest aesthetic evaluation [48].

Regarding contrast, Leder notes that a higher contrast is often perceived as increased familiarity, in turn leading to an affective preference (see following point) [42]. A related finding of his study is that clearer versions of a picture are preferred over less clear versions.

Another relevant aspect in Leder's model is familiarity, having been found to positively influence aesthetic preferences [42]. This is explained by the fact that the recognition of familiar visual elements or structures is perceived as a rewarding experience, triggering positive emotional responses.

Finally, Leder highlights the importance of the Gestalt laws of Wertheimer, further elucidated by Girwidz for the aesthetic perception of pictures [11, 42, 49]. Moshagen & Thielsch argue that adherence to the Gestalt laws facilitates easier and more efficient picture processing [45]. The specific Gestalt laws that significantly contribute to the aesthetic appeal are the following: The law of proximity describes the tendency to group elements that are placed closer together. The law of similarity states that elements sharing a greater resemblance are interpreted as more closely related.

The use of pictures in the specific setting of physics education[1] leads to several additional criteria, both affective and practical. First, research about motivation in STEM education informs that selecting specific contexts for a given topic can increase interest and other affective variables [50 - 52; see 53 for a review], as well as foster meaningful learning [54]. Interesting contexts for science learning were found in areas such as biology [10, 55], astronomy [2], and natural phenomena and scenes, as already stated above [47, 56]. In particular, nature was identified as interesting context for both female and male learners [10, 55, 57]. The present study will use this type of context.

Second, the implementation into a given teaching setting requires several practical criteria to be fulfilled: the topic of picture should be relevant for the teaching content and the learning objectives; it should depict the physics phenomenon in question in a clear and comprehensible manner; and it should have the potential to create a picture-related task to allow for the specific form of decorative-instructional pictures investigated in this study (specifically, optics in 7[th] and 8[th] grade). Another practical requirement is of course the right to use the picture in a given setting.

Table I gives an overview of the above criteria (aesthetic, affective, as well as practical) for the selection of pictures for the use in physics tasks.

*Table I: Criteria for the selection of aesthetic pictures for the use in physics tasks*

| Criteria | Background |
|---|---|
| **Aesthetic criteria** | |
| Color | Fundamental characteristic of aesthetic objects; general preference for more saturated colors, as long as natural tones are preserved. Lit: Leder et al. [42], Moshagen & Thielsch [45], Plass et al. [31], Joye et al. [47], Deng [44] |
| Visual complexity | Preference for a moderate level of visual complexity Lit: Berlyne [43], Leder et al. [42], Schmidt et al. [48] |
| Contrast | Preference for high contrast Lit: Leder et al. [42] |
| Familiarity | Recognition of familiar visual elements or structures is perceived as a rewarding experience, triggering positive emotional responses. Lit: Leder et al. [42] |
| Gestalt laws | Objects are not misleadingly near to other objects. Elements that share a greater resemblance are related. The law of proximity describes the tendency to group elements that are placed closer together. The law of similarity states that elements sharing a greater resemblance are interpreted as more closely related. Lit: Leder et al. [42], Wertheimer [49], Girwidz [11], Moshagen & Thielsch [45] |
| **Affective criteria** | |
| Context | Specific contexts can increase interest and other affective variables. Lit: Kaplan & Kaplan [56], Muckenfuß [54], Hoffmann [10], Joye et al. [47], Bennett [53] |
| **Practical criteria** | |
| Topic | Relevant for teaching content and learning objectives |
| Content | Represents the physics phenomenon in question in a clear and comprehensible manner; has the potential to create a picture-related task |
| Rights of use | Available and free for education and teaching |

### C. Research objectives

Based on the above research background, the present study pursues two research objectives:

Firstly, we applied the derived criteria to select aesthetic pictures in a given field of physics teaching (geometrical optics) and we developed and validated instruments to evaluate learners' actual aesthetic and affective perception of these selected pictures.

Secondly, a classroom study on the aesthetic and affective perception of the selected pictures using this instrument was conducted. Specifically, this study addresses the following research question: How does the aesthetic and affective perception of the selected aesthetic pictures (based on the specified criteria) in tasks by learners compare to the perception of conventional instructional pictures in tasks on the same physics content?

# III. Materials and methods

### A. Setting and study design

This study involved in total 118 participants from six classes at different German (Bavarian) lower secondary schools (International Standard Classification of Education (ISCED) level 2.4.4; [58]). Complete data sets are available for 111 participants for the test described in the following. 51% of the participants were female ($N = 57$), and the average age of the participants was 13.4 years ($SD = 0.7$). Due to ethical considerations, a crossover design was chosen. In a 30-min long intervention, participants completed interactive tasks with alternating aesthetic pictures (AP) and classroom experiment pictures (CEP), see Fig. 1. Each class was divided into two groups: Group 1 ($N_{total} = 55$, grey in Fig. 1) started with APs, while group 2 ($N_{total} = 56$, yellow frame in Fig. 1) began with CEPs. For each picture topic - light propagation/shadows, spectrum, reflection, and refraction – participants alternated between APs and CEPs. These topics are covered in this order in the 7th and 8th grade in Bavarian high schools. After completing their picture-related tasks (see Sec. III C for details), participants provided an evaluation of their aesthetic and affective perception for the pictures they had engaged with (see Sec. III D for details).

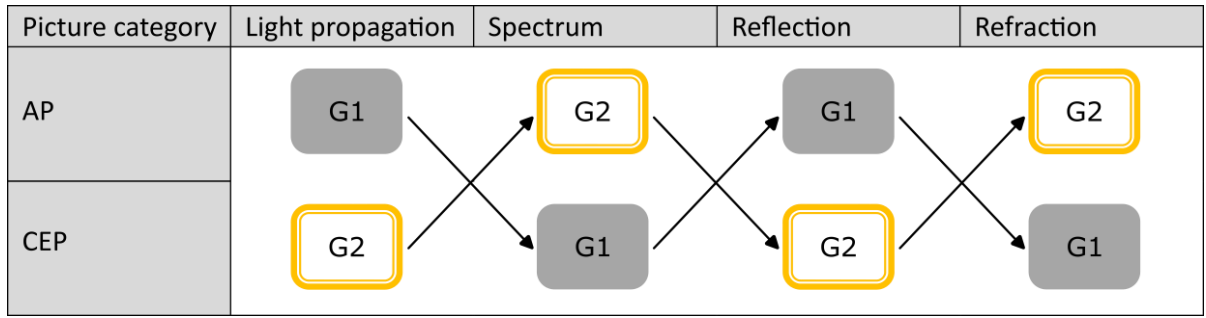

*Figure 1: Visualization of the crossover design.*

In the compact notation according to Shadish the design is summarized as follows [59]:
R   O   $X_1Y_1$   $X_2Y_2$   $X_1Y_3$   $X_2Y_4$   O
R   O   $X_2Y_1$   $X_1Y_2$   $X_2Y_3$   $X_1Y_4$   O
The abbreviations stand for:
X = independent variable 1, here the picture category; subscripts refer to AP or CEP
Y = independent variable 2, here the picture topic; subscripts refer to the four topics
O = observation
R = indication that participants have been randomly assigned

### B. Description of the selected pictures

Four essential concepts of geometrical optics were chosen in accordance with the local study plan: light propagation/shadows, spectrum, reflection, and refraction. We selected four APs that meet the aesthetic and the content-related criteria outlined in Sec. II B. These pictures depict a sunset, a rainbow in a landscape, a mountain with its reflection in a lake, and a landscape with its real image through a glass sphere, and thus cover the four concepts from geometrical optics. To contrast the APs with less aesthetic counterparts, we created photographs which depict the same physical phenomena, but with a classroom experiment as context (CEPs).

Table II provides a comparison between the two categories of pictures in line with the research background described above (Sec. II B; Tab. I) exemplified for the topic of reflection: APs are more colorful, have a higher, though still moderate level visual complexity, and show a more familiar context (a natural mountain scene) of higher motivational potential then the corresponding CEPs. Also, the created CEPs adhered to the principles of Gestalt laws. Both APs and CEPs meet the practical requirements and also have good contrast, as necessary in a teaching context.

Table III lists general differences and similarities between APs and CEPs and comments on the degree of realization of the criteria for the three other topics.

*Table II: Implementation of the criteria presented in Table I in an exemplary aesthetic picture (AP) and classroom experiment picture (CEP) about reflection.*

| Criteria | Realization in an exemplary AP | Realization in an exemplary CEP |
|---|---|---|
| | 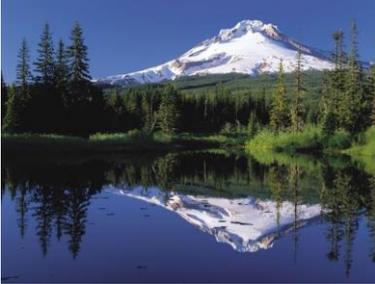 | 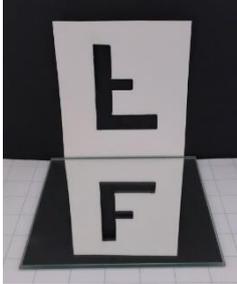 |
| **Aesthetic criteria** | | |
| Color | Color picture | Black and white picture |
| Visual complexity | Moderate: a mountain landscape with a forest and lake (plus mirror image) | Low: one letter on a cardboard (plus mirror image) |
| Contrast | Dark and bright colors alternate and yield a high contrast. | Black and white colors alternate and yield a high contrast. |
| Familiarity | Familiar type of landscape | Less familiar setting (physics experiment) |
| Gestalt laws | With the high color contrast, objects are either clear to distinguish from each other or are grouped together. | The high contrast makes objects clear to distinguish from each other. |
| **Affective and practical criteria** | | |
| Context | Nature as motivating context | Classroom experiment |
| Topic | Geometrical optics is part of the local curriculum in Bavaria, Germany; reflection is addressed in 7th and 8th grade. | |
| Content | All relevant objects for mirror reflection are visible: the mountain as the object, the lake as the optical mirror, and the reflected mountain as the optic image. | All relevant objects for mirror reflection are visible: the black letter F as the object, the small mirror as the optical mirror, and the reflected letter F as the optic image. |
| Rights of use | Public domain [60] | Created by author |

*Table III: Overview of the implementation of the criteria (AP vs. CEP) for the other topics.*

| | AP | CEP |
|---|---|---|
| **General differences** | Pictures are colorful and have nature as interesting and familiar context. Natural scenes are of higher complexity than for CEPs, and thus contain elements irrelevant for the physical concept/phenomenon in question. | Pictures are not colorful and have classroom experiments as less interesting and familiar context. They have lower complexity and the number of irrelevant objects is minimal. |
| **General similarities** | Both categories of pictures meet the practical requirements, and also have sufficient contrast for easy recognition of the main elements. | |

| **Light propagation/shadows** | 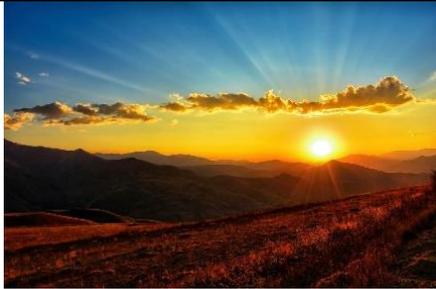 |
|---|---|
| Aesthetic criteria | See general differences. |
| Content | Both pictures show the essential optical elements: light source, obstacles, and shadows. |
| Other Criteria | See general similarities (picture sources: AP by [61][2], CEP by author). |

| **Spectrum** | 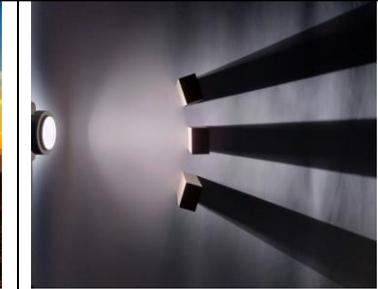 |
|---|---|
| Aesthetic criteria | See general differences, note however that in this case the CEP also fulfills the criterion of colorfulness. |
| Content | When photographing a rainbow in nature, it is not possible to frame the light sun as light source, rainfall (as dispersive medium) and the rainbow in the same time. | The picture shows the essential optical elements: light source, the dispersive medium and the spectrum. |
| Other Criteria | See general similarities (picture sources: AP by [62], CEP by author). |

| **Refraction** | 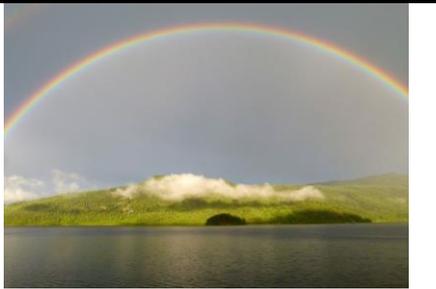 |
|---|---|
| Aesthetic criteria | See general differences. |
| Content | Both pictures show the essential optical elements: object, lens, and inverted image. Note, however, that in this case the mounting in the CEP are irrelevant elements for the optical phenomenon. |
| Other Criteria | See general similarities (picture sources: AP by [63] CEP by author). |

## C. Instructional design

In the present work, we investigate a specific approach combining the decorative and the instructional functions of pictures in tasks. Accordingly, an interactive learning environment was developed in which participants engaged with the pictures in a workbook-like format on tablets. Each of the four picture topics (see Sec. III B) included two pictures from the same category - either AP or CEP- and up to ten picture-related tasks in total. The first picture was presented at the top of the page, followed by tasks related to this picture and a second picture with related tasks. This structure ensured that par-ticipants actively interacted with the pictures and used them instructionally. The picture-related tasks across all picture topics adhered to a consistent framework:

- focus on and identify task relevant picture elements in the first AP
- link the identified picture elements to objects of the corresponding physics phenomenon
- recognize the physical phenomenon in a second AP
- summarize the physical phenomenon.

---

[2] The light propagation/shadows photograph is a derivative representation closely resembling the original depiction, which cannot be displayed due to restricted permissions.

A representative task related to aesthetic pictures is presented in Figure 2. This task required participants to link identified picture elements to objects of the corresponding physical phenomenon. The corresponding tasks related to CEPs were identical in content. After completing the picture-related tasks for all topics, participants were asked to rate the pictures in terms of aesthetic and affective picture perception (see next section).

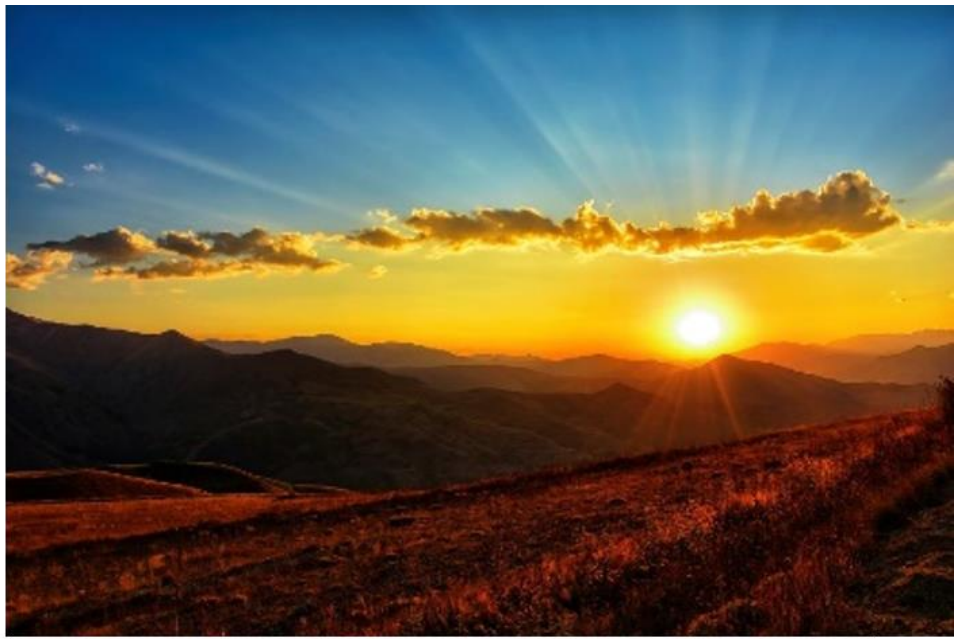

*Figure 2: Aesthetic picture task (topic light propagation/shadows). The picture-related task asks participants to relate objects in the picture with terms related to the physical phenomenon (e.g. clouds – opaque obstacle). Picture source: [61]*

## D. Instruments

The instrument for aesthetic and affective perception of pictures comprises two components.

The first component, the instrument for the aesthetic perception of pictures (IaePP), includes five items derived from a study of Miller [64], which is itself based on extensive research on perceived visual aesthetics by Lavie & Tractinsky [65]. Miller's subscale consists of ten items aimed at the aesthetic perception of websites [64]. His analyses supported combining these subscales into a single scale. For the IaePP, we selected items of both subscales, ensuring their suitability for the evaluation of photographs, and avoiding linguistic similarities after translation into German. Here, participants were asked to rate their perception of each picture in terms of 'creative', 'appealing', 'fascinating', 'original' and 'aesthetic'.

The second component, the instrument of the affective perception of pictures (IafPP), comprises five items. These items are derived from a validated instrument by Steyer et al., which measures three dimensions of affective state [66]: mood (good/bad), alertness (awake/sleepy), and calmness (calm/restless). The affective state instrument has already been used in research on decorative pictures [6]. Participants were asked to describe their mood for each picture in these dimensions.

All items of the IaePP and of the IafPP used a six-point rating scale (1 = lowest score, 6 = highest score). The original German wording and the English translation of the items are provided in the appendix (Table SI).

As a complement, we conducted short interviews to check for consistency of the quantitative results with qualitative information about participants' perception of the tested aesthetic pictures.

## E. Statistical analyses

The statistical analyses were carried out using the R software package, and Excel® [67]. In a preliminary data screening, datasets exhibiting distinct patterns like sequential (zigzag) response styles were excluded from the analysis, as these patterns indicated that participants were no longer actively engaging with the item texts.

After data screening, multiple statistical analyses were performed. Initially, to validate the instruments designed to measure the aesthetic and affective picture perception, several item characteristics were calculated according to standard procedures: Cronbach's α as measure of internal consistency, the mean discrimination index $\bar{D}$ and the mean of the item-test-correlation $\bar{r}_{it}$ [68, 69]. Following conventional psychometric standards, items were considered acceptable if they met the critical thresholds of $\alpha > 0.7$, $\bar{D} > 0.2$ and $\bar{r}_{it} > 0.3$ [70, 68]. Afterwards, the following more detailed statistical analyses were performed, in accordance with Field et al. [70]. First, the normality of the ratings for each picture was tested using the Shapiro-Wilk normality test for both IaePP and IafPP. Additionally, Levene's test indicated homogeneity of variances between AP and CEP for every picture topic ($p > 0.05$). Residuals were normally distributed.

Then, a robust two-way mixed ANOVA was conducted to examine the influence of the picture category (AP or CEP) and the picture topic (light propagation/shadows, spectrum, reflection, refraction) on the dependent variable (IaePP and IafPP). Analyses within each picture category were carried out by Welch's Two Sample $t$-tests. Finally, to confirm the validity of the analyses of the non-normal data, an additional non-parametric analysis was performed with the Wilcoxon rank sum test with continuity correction. These non-parametric tests results are used to verify the interpretation of non-normal data.

The predictor variables were analyzed using Welch's Two Sample $t$-tests for gender effects and Spearman's rank correlation coefficients for the effects of intrinsic motivation/engagement, interest in physics and interest in photography.

Effect sizes were calculated as Cohen's $d$ using the pooled standard deviation according to standard procedures ($d = (M_1 - M_2)/SD_p$, where the $M$s are the means of the compared groups, and $SD_p$ is the pooled standard deviation [71, 72]. For their corresponding confidence intervals, we used formulas described by Ben-Shachar et al. [73]. Conventional effect size levels used for the following discussion are small ($0.2 < d < 0.5$), medium ($0.5 \leq d < 0.8$) or large ($0.8 \leq d$) [71]. Another reference value of $d = 0.4$ is used by Hattie as an 'hinge point' between influences of smaller and larger size[3] [74].

# IV. Results

## A. Validation of instruments

The main psychometric indicators, i.e. internal consistency (Cronbach's $\alpha$), the mean of discrimination indices $\bar{D}$, and the mean of item-test correlations $\bar{r}_{it}$ for both instruments are presented in Table IV. For both instruments, the mean of the discrimination indices and the means of the corrected item-test correlations met the conventional critical values ($\bar{D} > 0.2$ and $\bar{r}_{it} > 0.3$, see in Sec. III E). Additionally, the internal consistency exceeded $\alpha = 0.7$ for both instruments.

*Table IV: Psychometric indicators of the instruments of aesthetic and affective picture perception: number of items ($N_I$), internal consistency (Cronbach's $\alpha$), mean of discrimination indices ($\bar{D}$) and mean of item-test correlations ($\bar{r}_{it}$).*

| Instrument | $N_I$ | $\alpha$ | $\bar{D}$ | $\bar{r}_{it}$ |
|---|---|---|---|---|
| IaePP | 5 | 0.87 [0.85 - 0.89] | 0.79 | 0.57 |
| IafPP | 5 | 0.82 [0.80 - 0.85] | 0.76 | 0.48 |

## B. Effects of the intervention

The robust two-way mixed ANOVAs indicated significant main effects of picture category (AP vs CEP) and picture topic (light propagation/shadows, spectrum, reflection, refraction). The main effects was the picture category with IaePP: $F(1, 108) = 65.55$, $p < 0.05$; and IafPP: $F(1, 108) = 62.71$, $p < 0.05$. The picture topic had notably lower values for IaePP: $F(3, 324) = 6.85$, $p < 0.05$; and IafPP: $F(3, 324) = 4.01$, $p < 0.05$. There were significant interaction effects between the picture category and the picture topic: $F(3, 324) = 10.38$, $p < 0.05$ for IaePP and $F(3, 324) = 7.54$, $p < 0.05$ for IafPP.

Due to the significant influence of picture topic, additional analyses were carried out for each topic individually. For all picture topics except 'spectrum', Welch's Two Sample $t$-tests revealed significant differences between APs and CEPs. The test statistics and effect sizes are reported in
Table V.

Effect sizes (Cohen's $d$) are provided in Fig. 3. Large values were found for both aesthetic ($d = 1.05$ – 1.56) and affective picture perception ($d = 0.85$ – 1.48) for the topics of light propagation/shadows, reflection and refraction. The only exception is the topic 'spectrum' with no significant differences for IaePP nor IafPP, which will be discussed below.

Additionally, a non-parametric Wilcoxon test confirmed the significant differences found between AP and CEP, and the effect sizes were within 10% of those calculated with the parametric test (see supplementary Table SII).

Supplementary qualitative data from short interviews supported these findings. Participants were asked what they liked or disliked about the pictures used in the study. Answers primarily felled into two kinds of criteria: First, criteria related to image features such as color, contrast and visual complexity, consistent with research in the psychology of perception of aesthetics and art taken as basis of the present research. The second kind of criteria is related to the content of visual context of the pictures such as 'sunset' and 'landscape'. The qualitative data from the interviews thus are consistent with the criteria of aesthetic and affective appeal used in the present study (Sec. II B).

---

[3] We agree with Hattie (2009) that these thresholds are an element of discussion to be used with circumspection, not values to be blindly applied.

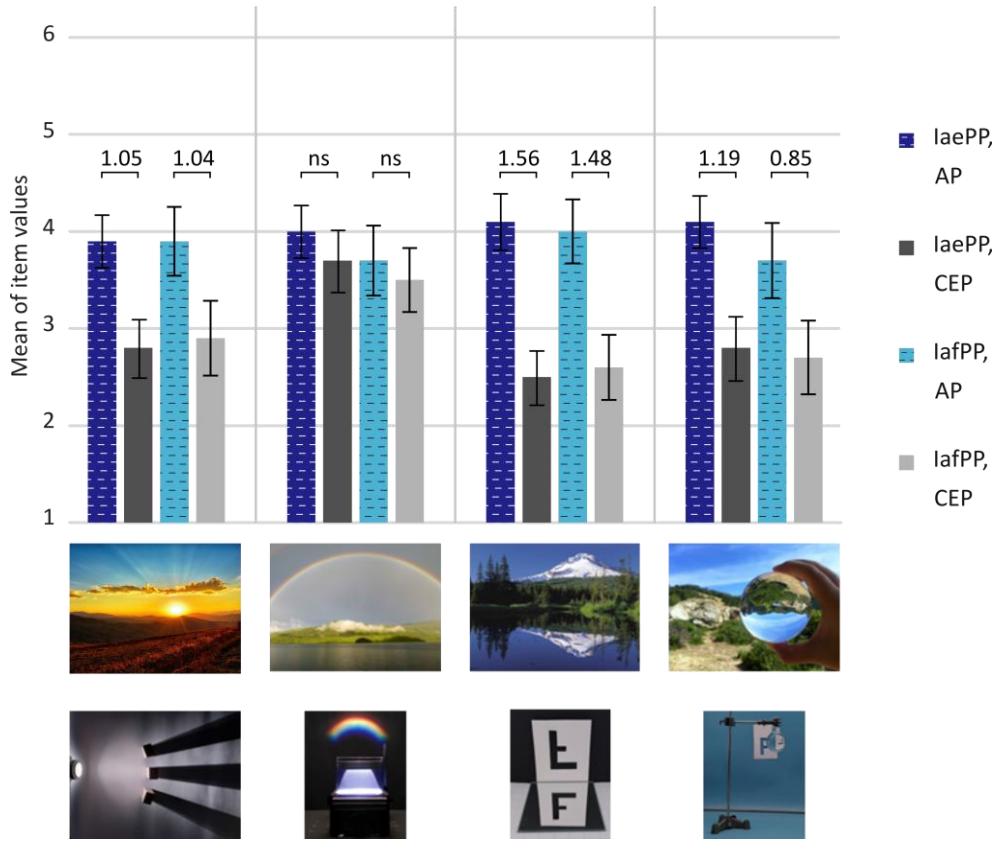

*Figure 3: Means with confidence intervals of the aesthetic and affective perception of APs and CEPs. The numbers on top of the brackets report Cohen's d values. The abbreviations are: AP, aesthetic picture; CEP, classroom experiment picture; IaePP, instrument for aesthetic perception of pictures; IafPP, instrument for affective perception of pictures.*

*Table V: Statistics of the aesthetic and affective perception of pictures regarding differences between APs and CEPs: means and SD as well as t, df, p and Cohen's d from Welch's Two Sample t-tests.*

| Picture Topic | Variable | Mean (*SD*) | | Welch's Two Sample *t*-test | | | |
|---|---|---|---|---|---|---|---|
| | | AP | CEP | *t* | *df* | *p* | *d* (CI) |
| Light propagation | IaePP | 3.9 (1.0) | 2.8 (1.1) | 5.52 | 108 | 0.00 | 1.05 (0.4) |
| | IafPP | 3.9 (0.8) | 2.9 (1.1) | 5.47 | 101 | 0.00 | 1.04 (0.4) |
| Spectrum | IaePP | 4.0 (1.0) | 3.7 (1.2) | 1.32 | 106 | 0.09 | ns |
| | IafPP | 3.7 (1.2) | 3.5 (0.8) | 1.08 | 98 | 0.14 | ns |
| Reflection | IaePP | 4.1 (1.1) | 2.5 (1.1) | 7.49 | 107 | 0.00 | 1.56 (0.43) |
| | IafPP | 4.0 (1.0) | 2.6 (1.0) | 7.74 | 108 | 0.00 | 1.48 (0.42) |
| Refraction | IaePP | 4.1 (1.0) | 2.8 (1.2) | 6.23 | 103 | 0.00 | 1.19 (0.41) |
| | IafPP | 3.7 (1.2) | 2.7 (1.1) | 4.46 | 108 | 0.00 | 0.85 (0.39) |

### C. Influence of potential predictor variables

Regarding gender, the rating of the APs by female participants (IaePP: $M = 4.23$, $SD = 0.74$, IafPP: $M = 4.01$, $SD = 0.87$) was higher than that of male participants (IaePP: $M = 3.77$, $SD = 1.24$, IafPP: $M = 3.63$, $SD = 1.16$). This difference turned out to be significant (IaePP: $t(173) = 3.32$, $p < 0.005$, IafPP: $t(199) = 2.80$, $p < 0.01$), with effect sizes around $d = 0.4$ (IaePP: $d = 0.45$, [0.18 - 0.72], IafPP: $d = 0.38$, [0.11 - 0.65]). There were no significant differences in the ratings of CEPs between female (IaePP: $M = 2.98$, $SD = 1.13$, IafPP: $M = 3.01$, $SD = 1.01$) and male (IaePP: $M = 2.83$, $SD = 1.30$, IafPP: $M = 2.83$, $SD = 1.13$) participants.

Regarding a possible influence of intrinsic motivation/engagement (IME), interest in physics (I_Phy) and interest in photography (I_Pho), the analysis revealed no significant correlations of the affective perception for any combination of picture categories with topics. For the aesthetic perception, the only two weak but statistically significant correlations were between IME and I_Phy and the rating of the APs: (IME: $\rho = 0.35$, 95% CI: [0.18, 0.41], $p < 0.001$; I_Phy: $\rho = 0.32$, 95% CI: [0.17, 0.43], $p < 0.001$).

# V. Discussion and conclusions

### A. Main outcomes

The purpose of the present work was to investigate the potential of aesthetic pictures as used in picture-related tasks for physics education. Specifically, the study aimed at two research objectives (Sec. II C), (i), the development of instruments about the aesthetic and affective perception of science-related pictures and (ii) a classroom study on the aesthetic and affective perception of aesthetic pictures in picture-related tasks as compared to the perception of conventional instructional pictures in tasks about the same physics content. We now discuss the findings of the study with respect to these research objectives.

Instrument to evaluate the aesthetic and affective picture perception (research objective 1)

The psychometric indicators of the instrument (discrimination, item-test-correlation, internal consistency) meet the standard quality criteria for both aesthetic (IaePP) and affective picture perception (IaePP) [68]. In particular, internal consistencies (IaePP: $\alpha_C = 0.87$ [0.85, 0.89]; IafPP: $\alpha_C = 0.82$ [0.80, 0.85]) are well above of the threshold for group level measurements of 0.7 recommended by the European Federation of Psychological Associations [75]. This might be considered as a non-trivial outcome for the investigation of variables with a high degree of subjective variability, and we conclude that the instrument with its two components can be useful for the aesthetic and affective evaluation of science-related pictures for research and teaching purposes.

The observed aesthetic and affective perception of pictures in tasks (research objective 2)

Results revealed a significantly better evaluation of the aesthetic pictures in tasks, selected according to the specified criteria, with large effect sizes compared to pictures depicting classroom experiments in tasks covering the same physics content (AP vs. CEP, aesthetic and affective perception: $d = 1.05 - 1.56$ and $0.85 - 1.48$, respectively). The only exception to this was the picture pair related to the spectrum. While the rating for the AP was comparable to that of other picture topics, the rating of this particular CEP was higher than that of the other CEPs, and no statistically significant difference to the AP counterpart was found. This, however, can be explained by the colorfulness of this specific CEP, inherent to the topic of 'spectrum'.

These findings thus validate the criteria for the selection of aesthetic pictures as derived here from prior research, and we conclude that the procedure presented in this study allows to effectively select aesthetically and affectively appealing pictures for research and teaching purposes in physics education.[1]

### B. Limitations and specificity of the results

Several limitations have to be mentioned. The present study has investigated photographs of natural scenes as aesthetic pictures for tasks about geometrical optics, which is a specific choice for the technical picture type, for the aesthetic context, and for the teaching subject. As for other picture types, artificially generated illustrations could be considered, especially for the relatively ease to manipulate factors such as hue, visual complexity, contrast, and other. Such artificial pictures ("artist's view") as widespread e.g. in science popularization media can indeed convey an aesthetical impression, and with the advent of advanced AI tools like GPT-4 they are easier and easier to craft. This could e.g. also allow for contexts customized to specific interests of participants, a setting of potential interest both for practice and research. On the other hand, care must be taken to ensure that these advantages of artificial pictures do not turn into a disadvantage. Viewers are often exposed to potential manipulation and misinformation by such visualizations in their daily lives. As a result, the trust regarding the accuracy of such artificial pictures can be lower, and whether this leads to lower engagement in physics tasks related to them remains an open research question.

Furthermore, the approach of this study can be expanded to pictures with aesthetic contexts beyond natural scenes visible with the naked eye, such as micrographs or astronomical pictures. The aesthetic potential of such pictures is well documented e.g. by international competitions and awards for microscopic pictures (e.g. Nikon Small World [76]), or by the existence of a market for large-format picture books on astronomy (e.g. about the Hubble Telescope [77]). Finally, the approach presented here could be extended to other teaching topics like mechanics (e.g. tasks using time lapse photography) or thermal physics (tasks using thermography).

On the theoretical level, no claim is made about the completeness of the set of criteria used in this study. Although research-based and sufficient for selecting pictures that are perceived as aesthetic in a teaching context (i.e., when used in tasks), aesthetic evaluation is a very broad field, and other relevant criteria may exist that could further improve the specification and selection of aesthetic pictures for educational purposes.

### C. Research perspectives

An interesting direction for future research would be to consider two methodological aspects not covered in the present study. The first aspect addresses the 'disentanglement' of criteria, meaning a more detailed investigation into the relative strength of the different factors contributing to the aesthetic quality of pictures. The comparison between the AP and CEP concerning the pictures about the spectrum of visual light suggests that color might be a stronger factor than context in this case. A quantification of the strength of the various influences could also provide hints for the question raised above, i.e. whether there are other influential factors to be included in criteria set. To explore this further, quantitative studies with larger picture sets and greater variation in criterion factors would be necessary, along with more in-depth qualitative studies on viewer perceptions. Second, beyond a focus on individual criteria, several types of interaction might be of interest, e.g. between aesthetic and affective criteria, or between criteria and viewer characteristics (e.g. degree of expertise). Models for such a more complex and complete analysis have been suggested [42, 47], but this is beyond the scope of the present work.

Finally, an interesting aspect for future research would be to examine also the impact of tasks involving aesthetic pictures on cognitive variables, such as cognitive engagement, and ultimately, learning.

### D. Teaching perspectives and concluding remark

From our perspective, a main outcome of this study relevant to teaching practice is to overcome the long-standing dichotomy between decorative and instructional pictures. Aesthetic pictures, when integrated into tasks, can effectively serve an instructional purpose beyond a merely decorative function with its distractive side-effects.

Aesthetic pictures can thus be used to enrich physics teaching (at least for motivational purposes). Regarding practical implementation, we see two stages in the learning process in which they would fit. Tasks with aesthetic pictures (APs), such as classroom questions, can be used at the beginning of a lesson or teaching sequence to provide immersion in a topic and spark interest for further engagement with it. In a later stage, APs can be applied in tasks for application and transfer.

We see the development of a full-fledged instructional sequence that exploits this potential, accompanied by a study addressing relevant additional research questions, particularly concerning learning outcomes, as a useful and interesting next step based on the present study.

**Author Contributions:** Conceptualization: T.L., R.G., A.M.; Formal analyses T.L., R.G., A.M.; Funding acquisition: A.M.; Investigation T.L.; Methodology: T.L., R.G., A.M.; Writing - draft T.L. and R.G.; Writing - review and editing T.L., R.G., A.M.

**Acknowledgements:** T.L. acknowledges funding by the Wilfried und Ingrid Kuhn foundation. Open Access funding is enabled and organized by LMU.

# Supplementary material

**Table SI:** The following table presents the introduction to the aesthetic and affective perception of pictures as well as the item texts. Item texts are in its original language German as well as translated into English. IaePP refers to items of the instrument for aesthetic picture perception, while IaePP refers to items of the instrument for affective perception of pictures.

|  | Introduction, translated into English | Introduction, original text |
|---|---|---|
| General introduction to this part of the questionnaire | Finally, please rate the pictures. There is no right or wrong here, just follow your intuition! There are also a few questions related to physics. | Zum Schluss bewerte bitte die verwendeten Bilder. Hier gibt es kein richtig und falsch, folge einfach deiner Intuition! Außerdem gibt es noch ein paar fachliche Fragen. |
| Introduction to each picture | Please give your personal assessment of the picture shown below. There is no right and no wrong. | Im Folgenden gebe bitte deine persönliche Einschätzung zu dem hier gezeigten Bild an. Es gibt kein Richtig und kein Falsch. |
| Item no. | Item translation into English | Item original text |
| IaePP_1 | I find the picture aesthetic. | Das Bild finde ich ästhetisch. |
| IaePP_2 | I find the picture appealing. | Das Bild finde ich ansprechend. |
| IaePP_4 | I find the picture creative. | Das Bild finde ich kreativ. |
| IaePP_5 | I find the picture fascinating. | Das Bild finde ich faszinierend. |
| IaePP_6 | I find the picture original. | Das Bild finde ich originell. |
| IafPP_1 | While looking at the picture, I feel happy. | Beim Betrachten des Bildes fühle ich mich glücklich. |
| IafPP_2 | While looking at the picture, I feel bored. | Beim Betrachten des Bildes fühle ich mich gelangweilt. |
| IafPP_4 | While looking at the picture, I feel good. | Beim Betrachten des Bildes fühle ich mich gut. |
| IafPP_5 | While looking at the picture, I feel awake. | Beim Betrachten des Bildes fühle ich mich wach. |
| IafPP_6 | While looking at the picture, I feel calm. | Beim Betrachten des Bildes fühle ich mich ruhig. |

**Table SII:** Statistics of the aesthetic and affective perception of pictures (IaePP and IafPP) regarding differences between the aesthetic pictures (AP) and classroom experiment pictures (CEP). The table shows means and *SD* as well as *t*, *df*, *p* and Cohen's *d* from Welch's Two Sample *t*-tests. Additionally, the results for the nonparametric Wilcoxon rank sum tests with continuity correction are provided (Wilcoxon-W, *p*, effect sizes *r* and effect sizes with continuity correction *d*).

| Picture Topic | Variable | Mean (*SD*) | | *t*-test | | | | Wilcoxon rank sum test | | | |
|---|---|---|---|---|---|---|---|---|---|---|---|
|  |  | AP | CEP | *t* | *df* | *p* | *d* (CI) | *W* | *p* | *r* | *d* |
| Light propagation | IaePP | 3.9 (1.0) | 2.8 (1.1) | 5.52 | 108 | 0.00 | 1.05 (0.4) | 2361 | 0.00 | 0.47 | 1.11 |
| | IafPP | 3.9 (0.8) | 2.9 (1.1) | 5.47 | 101 | 0.00 | 1.04 (0.4) | 2329 | 0.00 | 0.45 | 1.05 |
| Spectrum | IaePP | 4.0 (1.0) | 3.7 (1.2) | 1.32 | 106 | 0.09 | 0.25 (0.38) | 1767 | 0.13 | 0.12 | ns |
| | IafPP | 3.7 (1.2) | 3.5 (0.8) | 1.08 | 98 | 0.14 | 0.21 (0.37) | 1792 | 0.10 | 0.16 | ns |
| Reflection | IaePP | 4.1 (1.1) | 2.5 (1.1) | 7.49 | 107 | 0.00 | 1.56 (0.43) | 2641 | 0.00 | 0.61 | 1.68 |
| | IafPP | 4.0 (1.0) | 2.6 (1.0) | 7.74 | 108 | 0.00 | 1.48 (0.42) | 2606 | 0.00 | 0.59 | 1.60 |
| Refraction | IaePP | 4.1 (1.0) | 2.8 (1.2) | 6.23 | 103 | 0.00 | 1.19 (0.41) | 2432 | 0.00 | 0.53 | 1.23 |
| | IafPP | 3.7 (1.2) | 2.7 (1.1) | 4.46 | 108 | 0.00 | 0.85 (0.39) | 2248 | 0.00 | 0.43 | 0.92 |

# References

1. Girwidz, R. (2009). Bilder und bildhafte Darstellungen: visuelle Darstellungsmittel im Unterricht nutzen. *Naturwissenschaften im Unterricht Physik*, 109

2. Smith, L. F., Smith, J. K., Arcand, K. K., Smith, R. K., Bookbinder, J. & Keach, K. (2011). Aesthetics and Astronomy: Studying the Public's Perception and Understanding of Imagery From Space. *Science Communication, 33*(2), 201–238. https://doi.org/10.1177/1075547010379579

3. Cooper, J. L., Sidney, P. G. & Alibali, M. W. (2018). Who Benefits from Diagrams and Illustrations in Math Problems? Ability and Attitudes Matter. *Applied Cognitive Psychology*, *32*(1), 24–38. https://doi.org/10.1002/acp.3371

4. Robin, H. (1992). *The scientific image from cave to computer*. New York: Harry N. Abrams.

5. Lindner, M. A. (2020). Representational and decorative pictures in science and mathematics tests: Do they make a difference? *Learning and Instruction, 68*(101345), 1–11. https://doi.org/10.1016/j.learninstruc.2020.101345

6. Lenzner, A., Schnotz, W. & Müller, A. (2013). The role of decorative pictures in learning. *Instructional Science, 41*(5), 811–831. https://doi.org/10.1007/s11251-012-9256-z

7. Schneider, S., Nebel, S. & Rey, G. D. (2016). Decorative pictures and emotional design in multimedia learning. *Learning and Instruction, 44*, 65–73. https://doi.org/10.1016/j.learninstruc.2016.03.002

8. Scharinger, C. (2023). Effects of emotional decorative pictures on cognitive load as assessed by pupil dilation and EEG frequency band power. *Applied Cognitive Psychology*, *37*(4), 861–875. https://doi.org/10.1002/acp.4087

9. Takahashi, S. (1995). Aesthetic properties of pictorial perception. *Psychological review, 102*(4), 671.

10. Hoffmann, L. (2002). Promoting girls' interest and achievement in physics classes for beginners. Learning and instruction, 12(4), 447-465.

11. Girwidz, R. (2020). Medien im Physikunterricht. In E. Kircher, R. Girwidz & H. E. Fischer (Eds.), *Physikdidaktik / Grundlagen,* (pp. 293–335). Springer Spektrum. https://doi.org/10.1007/978-3-662-59490-2_8

12. Peterson, M., Delgado, C., Tang, K. S., Bordas, C., & Norville, K. (2021). A taxonomy of cognitive image functions for science curriculum materials: identifying and creating 'performative'visual displays. *International Journal of Science Education*, 43(2), 314-343.

13. Carney, R. N. & Levin, J. R. (2002). Pictorial Illustrations Still Improve Students' Learning from Text. *Educational Psychology Review*, 14(1), 5–26. https://doi.org/10.1023/A:1013176309260

14. Renkl, A., & Scheiter, K. (2017). Studying visual displays: How to instructionally support learning. *Educational Psychology Review*, 29(3), 599-621. https://doi.org/10.1007/s10648-015-9340-4

15. Guo, D., McTigue, E. M., Matthews, S. D., & Zimmer, W. (2020). The impact of visual displays on learning across the disciplines: A systematic review. *Educational Psychology Review*, 32(3), 627-656.

16. Gilbert, J., Reiner, M. & Nakhleh, M. (2008). *Visualization theory and practice in science education*. New York: Springer.

17. Kim, M., & Jin, Q. (2022). Studies on visualisation in science classrooms: a systematic literature review. *International Journal of Science Education, 44*(17), 2613-2631.

18. Gilbert, J., & Treagust, D. F. (eds.). (2009). *Multiple Representations in Chemical Education*. Springer

19. Tsui, C. & Treagust, D. F. (eds.). (2013). *Multiple Representations in Biological Education*. Springer

20. Treagust, D. F., Duit, R., & Fischer, H. E. (eds.). (2017). *Multiple Representations in Physics Education*. Springer

21. Ploetzner, R., & Lowe, R. (2012). A systematic characterisation of expository animations. *Computers in Human Behavior, 28*(3), 781-794.

22. Girwidz, R. & Kohnle, A. (2022). Multimedia and digital media in physics instruction, in *Physics Education*, eds. H.E. Fischer, R. Girwidz. Cham : Springer International Publishing, pp. 297–336

23. Comenius, J. A. (1658, 1992). *Orbis sensualium pictus* (*The Visible World in Pictures)*; 1st ed., Nuremberg; engl. edition: Zürich: Pestalozzianum-Verlag / Comenius Verlag.

24. Levin, J. R. (1989). A transfer-appropriate-processing perspective of pictures in prose. In H. Mandl & J. R. Levin (Eds.), Knowledge acquisition from text and pictures (pp. 83-100). Amsterdam: Elsevier

25. Harp, S. F., & Mayer, R. E. (1997). The role of interest in learning from scientific text and illustrations: On the distinction between emotional interest and cognitive interest. *Journal of educational psychology*, 89(1), 92. https://doi.org/10.1037//0022-0663.89.1.92

26. Easterbrook, J. A. (1959). The effect of emotion on cue utilization and the organization of behavior. *Psychological Review, 66*(3), 183–201. https://doi.org/10.1037/h0047707

27. Ellis, H. C., & Ashbrook, P. W. (1988). Resource allocation model of the effects of depressed mood states. In K. Fiedler & J. Forgas (Eds.), *Affect, cognition and social behavior*. Toronto: Hogrefe.

28. Knörzer, L., Brünken, R. & Park, B. (2016). Facilitators or suppressors: Effects of experimentally induced emotions on multimedia learning. Learning and Instruction, 44, 97–107. https://doi.org/10.1016/j.learninstruc.2016.04.002

29. Mayer, R. E. (2005). Cognitive theory of multimedia learning. *The Cambridge handbook of multimedia learning, 41*, 31-48.

30. Rey, G. D. (2012). A review of research and a meta-analysis of the seductive detail effect. *Educational Research Review, 7*(3), 216-237. https://doi.org/10.1016/j.edurev.2012.05.003

31. Plass, J. L., Heidig, S., Hayward, E. O., Homer, B. D. & Um, E. (2014). Emotional design in multimedia learning: Effects of shape and color on affect and learning. *Learning and Instruction, 29*, 128–140. https://doi.org/10.1016/j.learninstruc.2013.02.006

32. Plass, J. L. & Kaplan, U. (2016). Emotional Design in Digital Media for Learning. In S. Y. Tettegah & M. Gartmeier (Hrsg.), *Emotions, Technology, Design, and Learning* (S. 131–161). Elsevier. https://doi.org/10.1016/B978-0-12-801856-9.00007-4

33. Abele-Brehm, A. (1992). Positive and negative mood influences on creativity: Evidence for asymmetrical effects. *Polish Psychological Bulletin, 23*(3), 203–221.

34. Schwarz, N. (1990). Feelings as information: Informational and motivational functions of affective states. In E. T. Higgins & R. M. Sorrentino (Eds.), *Handbook of motivation and cognition: Foundations of social behavior*, Vol. 2, pp. 527–561). The Guilford Press.

35. Schneider, S., Dyrna, J., Meier, L., Beege, M., & Rey, G. D. (2018). How affective charge and text–picture connectedness moderate the impact of decorative pictures on multimedia learning. *Journal of Educational Psychology, 110*(2), 233–249. https://doi.org/10.1037/edu0000209

36. Plass, J. L., & Kalyuga, S. (2019). Four ways of considering emotion in cognitive load theory. *Educational Psychology Review*, *31*, 339-359.

37. Schneider, S. (2017). The impact of decorative pictures on learning with media. Dissertation, Chemnitz: Technische Universität [Technical University]

38. Moreno, R. (2006). Does the modality principle hold for different media? A test of the method-affects-learning hypothesis. *Journal of Computer Assisted Learning, 22*(3), 149–158. https://doi.org/10.1111/j.1365-2729.2006.00170.x

39. Moreno, R. (2007). Optimizing learning from animations by minimizing cognitive load: Cognitive and affective consequences of signalling and segmentation methods. *Applied Cognitive Psychology, 21,* 765–781. https://doi.org/10.1002/acp.1348

40. Moreno, R., & Mayer, R. (2007). Interactive multimodal learning environments. Educational Psychology Review, 19(3), 309–326.

41. Scherer, D., Verkühlen, A., & Dutke, S. (2023). Effects of related decorative pictures on learning and metacognition. *Instructional Science*, *51*(4), 571-594. https://doi.org/10.1007/s11251-023-09618-8

42. Leder, H., Belke, B., Oeberst, A. & Augustin, D. (2004). A model of aesthetic appreciation and aesthetic judgments. *British journal of psychology (London, England: 1953), 95*(4), 489–508. https://doi.org/10.1348/0007126042369811

43. Berlyne, D. E. (1974). *Studies in the new experimental aesthetics: Steps toward an objective psychology of aesthetic appreciation.* Washington, DC: Hemisphere Publishing Corporation.

44. Deng, Y., Loy, C. C. & Tang, X. (2017). Image Aesthetic Assessment: An Experimental Survey. *IEEE Signal Processing Magazine, 34*(4), 80–106. https://doi.org/10.1109/MSP.2017.2696576

45. Moshagen, M. & Thielsch, M. T. (2010). Facets of visual aesthetics. *International Journal of Human-Computer Studies, 68*(10), 689–709. https://doi.org/10.1016/j.ijhcs.2010.05.006

46. Leder, H., Hakala, J., Peltoketo, V.T., Valuch, C. & Pelowski, M. (2022). Swipes and Saves: A Taxonomy of Factors Influencing Aesthetic Assessments and Perceived Beauty of Mobile Phone Photographs. *Frontiers in psychology, 13*, 786977.

47. Joye, Y., Pals, R., Steg, L. & Lewis-Evans, B. (2013). New methods for assessing the fascinating nature of nature experiences. *PloS one, 9*(1), e65332. https://doi.org/10.1371/journal.pone.0065332

48. Schmidt, T., & Wolff, C. (2018). The influence of user interface attributes on aesthetics. i-com, 17(1), 41-55.

49. Wertheimer, M. (1923). Untersuchungen zur Lehre von der Gestalt. II. *Psychologische Forschung, 4*(1), 301–350. https://doi.org/10.1007/bf00410640

50. Fensham, P. J. (2009). Real world contexts in PISA science: Implications for context-based science education. *Journal of Research in Science Teaching 46*, 884–896.

51. Kuhn, J., Müller, A. (2014). "Context-based Science Education by Newspaper Story Problems: A Study on Motivation and Learning Effects". Progress in Science Education 1, 5-21.

52. Aydin-Ceran, S. (2021). Contextual learning and teaching approach in 21st century science education. *Current Studies in Social Sciences*, 160-173.

53. Bennett, J. (2016). Bringing Science to Life - Research Evidence. In: Taconis, R., den Brok, P., Pilot, A. (Eds.). (2016). Teachers creating context-based learning environments in science. Rotterdam: Sense Publishers

54. Muckenfuß, H. (1995). *Lernen im sinnstiftenden Kontext: Entwurf einer zeitgemäßen Didaktik des Physikunterrichts.* Berlin: Cornelsen.

55. Hoffmann, L., Häussler, P. & Lehrke, M. (1998). *Die IPN-Interessenstudie Physik. IPN: Bd. 158.* IPN.

56. Kaplan, R. & Kaplan, S. (1989). *The experience of nature: A Psychological Perspective.* Cambridge University Press.

57. Labudde, P., & Metzger, S. (Eds.). (2019). *Fachdidaktik Naturwissenschaft. 1.-9. Schuljahr.* utb GmbH. https://doi.org/10.36198/9783838552071

58. Schneider, S. L. (2013). The International Standard Classification of Education 2011. In G. E. Birkelund (Hrsg.), *Comparative Social Research. Class and Stratification Analysis* (Bd. 30, S. 365–379). Emerald Group Publishing Limited. https://doi.org/10.1108/S0195-6310(2013)0000030017

59. Shadish, W. R., Cook, T. D. & Campbell, D. T. (2002). *Experimental and quasi-experimental designs for generalized causal inference* (Bd. 1195). Houghton Mifflin. https://iaes.cgiar.org/sites/default/files/pdf/147.pdf

60. Oregon's Mt. Hood Territory (2006). *Mount Hood reflected in Mirror Lake, Oregon* [Photograph] Wikimedia.
https://commons.wikimedia.org/wiki/File:Mount_Hood_reflected_in_Mirror_Lake,_Oregon.jpg

61. Bozkuş (12. April 2018). *Sunset-Dawn-Nature-Mountains* [Photograph] Pixabay. https://pixabay.com/photos/sunset-dawn-nature-mountains-3314275/

62. Werner (3. Juli 2013). *Regenbogen Canim See* [Photograph] Pixabay. https://pixabay.com/de/photos/regenbogen-canim-see-142701/

63. Gou (2016). *pꞁɹoʍ ǝɥʇ in the world* [Photograph] American Association of Physics Teachers. https://www.aapt.org/photodirectory/2016_C-731_in%20the%20world_lg.jpg

64. Miller, C. (2011). Aesthetics and e-assessment: the interplay of emotional design and learner performance. *Distance Education, 32*(3), 307–337. https://doi.org/10.1080/01587919.2011.610291

65. Lavie, T. & Tractinsky, N. (2004). Assessing dimensions of perceived visual aesthetics of web sites. *International Journal of Human-Computer Studies, 60*(3), 269–298. https://doi.org/10.1016/j.ijhcs.2003.09.002

66. Steyer, R., Schwenkmezger, P., Notz, P. & Eid, M. (1997). *Mehrdimensionaler Befindlichkeitsfragebogen (MDBF)*. Hogrefe. https://doi.org/10.1037/t12446-000

67. R Core Team (2022). R: A Language and environment for statistical computing. (Version 4.2.2) [Computer software]. Retrieved from https://cran.r-project.org

68. Ding, L. & Beichner, R. (2009). Approaches to data analysis of multiple-choice questions. Physical Review Special Topics - Physics Education Research, 5(2). https://doi.org/10.1103/PhysRevSTPER.5.020103

69. Panjaitan, R. L., Irawati, R., Sujana, A., Hanifah, N. & Djuanda, D. (2018). Item validity vs. item discrimination index: a redundancy? Journal of Physics: Conference Series, 983, 12101. https://doi.org/10.1088/1742-6596/983/1/012101

70. Field, A., Miles, J. & Field, Z. (2012). Discovering statistics using R. SAGE Publications Ltd.

71. Cohen, J. (1988). Statistical power analysis for the behavioral sciences (2nd ed.). Lawrence Erlbaum.

72. Fritz, C. O., Morris, P. E., & Richler, J. J. (2012). Effect size estimates: current use, calculations, and interpretation. Journal of experimental psychology: General, 141(1), 2.

73. Ben-Shachar, M., Lüdecke, D. & Makowski, D. (2020). effectsize: Estimation of Effect Size Indices and Standardized Parameters. Journal of Open Source Software, 5(56), 2815. https://doi.org/10.21105/joss.02815

74. Hattie, J. (2009). Visible learning: A synthesis of over 800 meta-analyses relating to achievement. New York: Routledge.

75. Evers, A., Hagemeister, C., & Hostmaelingen, A. (2013). EFPA Review Model for the description and evaluation of psychological and educational tests. Tech. Rep. Version 4.2. 6). Brussels: European Federation of Psychology Associations.

76. Nikon Small World, https://www.nikonsmallworld.com/, accessed on 19/9/2024

77. Bolden, C. F., Edwards, O., Grunsfeld, J. M., & Levay, Z. (2022). *Expanding universe: photographs from the Hubble space telescope*. Taschen.